\documentstyle[11pt,paspconf,epsf]{article}

\begin{document}

\title{The Bar-Halo Interaction in SB Galaxies}
\author{Victor P. Debattista and J.A. Sellwood}
\affil{Department of Physics and Astronomy, \\
Rutgers University, P.O. Box 849, Piscataway, NJ  08855}

\begin{abstract} \tolerance 5000
We describe fully self-consistent N-body experiments of barred galaxies with massive halos.  
A rotating bar is braked through dynamical friction with the halo, which occurs on a short time scale when the central density of the halo is high.   
On the other hand, friction is weak in a model with a central halo density low enough that the disk dominates the rotation curve in the inner parts; this model supports a fast bar for a Hubble time.  
We conclude that real barred galaxies in which the bar rotates rapidly, which is believed to be the rule, must have close to maximum disks.  
\end{abstract}

\section{Introduction}

The presence of a strongly triaxial object, a bar, at the center of some $30\%$ of disk galaxies provides an opportunity to probe the distribution of dark matter in the inner few disk scale lengths
and to test the maximum disk hypothesis.  
The bar pattern speed, $\Omega_p$, which plays an important role in these dynamics, can be parametrized by the ratio $s = D_L/a_B$.  
Here, $D_L$ is the distance from the center to the Lagrange point along the bar major axis (loosely known as corotation) and $a_B$ is the bar semi-major axis.  
Contopoulos (1980) argued that self-consistent bars can extend no further than co-rotation, i.e. $s \ge 1$.
For our purposes, we will consider bars to be fast when $1.0 \leq s \leq 1.4$.

The prevailing theory of bar formation requires a resonant cavity of spiral density waves (Toomre 1981). 
The cavity extends from the center to corotation, where spiral waves are reflected.  
Strong amplification at corotation leads to an instability that gives rise to a strong bar when it saturates.
Numerical simulations (e.g. Sellwood 1981; Combes and Sanders 1981) of unstable disks have revealed $s \simeq 1$ for bars formed in this way and those formed in our simulations also start out with a value not much larger than unity.

On the observational side, the evidence is meager but seems to indicate that bars are fast.  
Tremaine and Weinberg (1984) devised a direct method for measuring $\Omega_p$ which requires a tracer population that satisfies the continuity equation.  
It can therefore be used only for stellar populations free of obscuration, such as in SB0 galaxies.  
Kent (1987) and, more reliably, Merrifield and Kuijken (1995) have applied this method to NGC 936, finding $s = 1.4 \pm 0.3$, consistent with a fast bar.
All other determinations of bar pattern speeds are indirect. 
In particular, Athanassoula (1992) has modelled gas dynamics in barred galaxies. 
Identifing the bar dust lanes with gas shocks, she searched for a
value of $s$ which yielded shock patterns that match the observed dust
lane morphology, and concluded that $\Omega_p$ must be such that $s = 1.2 \pm 0.2$.  
Similar studies have been carried out by others, most notably using
potentials obtained from IR photometry (e.g. Lindblad et al. 1996, Weiner et al. 1996).  This evidence, which spans Hubble types SB0 to SBc, strongly suggests that all bars are fast in the sense defined above.  

However, the fact that all bars appear to rotate rapidly suggest that dynamical friction is weak. 
Chandrasekhar (1943) showed that a massive object moving through a uniform background of particles experiences a drag from the trailing wake that it induces.  
This retarding force is called dynamical friction.  
A similar process takes place when a bar rotates inside a halo: the bar excites a trailing, bisymmetric wake in the halo, which exerts a retarding torque on the bar.  
This torque transfers angular momentum from the bar to the halo, slowing the bar down.  
Linear perturbative calculations by Weinberg (1985) indicate that bars are slowed down in only a few rotations for Weinberg's massive halos.  
One then expects old bars inside massive halos to be slow.

\section{N-body Experiments}

Weinberg's calculations suffered from a number of shortcomings.
In particular, Weinberg modelled his bars with fixed analytic density distributions.  
Thus his treatment was not self-consistent and precluded any possibility of a back reaction on the bar (apart from slow-down).  
For example, Kormendy's (1979) hypothesis that a bar would dissolve could not be tested in Weinberg's work.
For this reason, we found it desirable to carry out {\it fully self-consistent} N-body simulations to test Weinberg's prediction.
The results of these simulations are presented below. 
Athanassoula (1996) is carrying out similar experiments. 

\subsection{Setup}

We have sought to generate galaxy models in which the halo is
initially in equilibrium with the disk.  We have accomplished this by
integrating iteratively the distribution function of the halo in the
potential of the halo and fixed disk until the halo density converges.  
We used either Kuz'min-Toomre or exponential disks, which were thicked vertically by a Gaussian factor.  
The vertical velocity dispersion in the disk was chosen to maintain this disk thickening.  
The disk velocity distribution in the radial direction was chosen to give a constant Toomre $Q$ at all radii, and we have experimented with different values.
Our halo distribution functions were lowered polytropes with different values of m:
\begin{equation}
f = f[(-E)^m] - f[(-E_0)^m] 
\end{equation}
where $E_0$ is the energy at some truncation radius.  
Quiet starts (e.g. Sellwood 1983) were used to suppress all components of the total linear momentum, and the disk-plane components of the total angular momentum. 
The system generated in this way is initially in equilibrium, but the
disk is not stable and a bar rapidly forms.
The simulations were run on 3D cartesian and polar grids; the grid codes used are described in Sellwood \& Merritt (1994) for the cartesian code and Sellwood \& Valluri (1996) for the polar code.  
We use units $G = M_d(\infty)/f = a = 1$, where $M_d(\infty)$ is the mass of the analytic disk model out to infinity, $f$ is the fraction of mass in the disk, and $a$ is the length scale of the disk.  
The unit of time is the dynamical time $t_{dyn} = \sqrt{a^3/GM}$.  
We parametrize the ratio of the halo mass to disk mass in the inner regions by the ratio $ \eta \equiv ( v_{c,disk}/v_{c,halo} )^2$ at the radius of the maximum total circular velocity in the plane of the disk.

\subsection{Massive halo model}

\begin{figure}[ht]
\plotfiddle{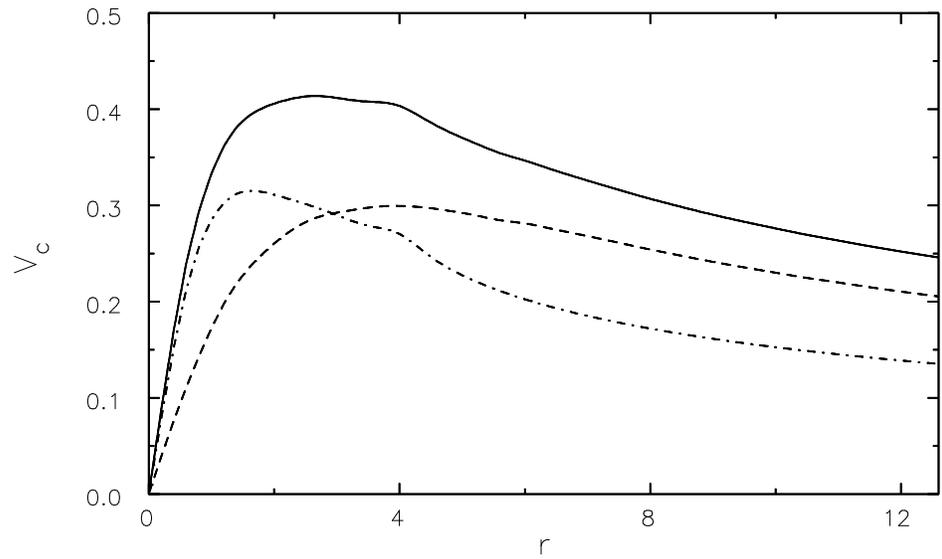}{4.0in}{0}{100}{100}{-290}{-270}
\caption[]{The circular velocity and its disk-halo decomposition for the
MH model.  The continuous line is the total rotation curve,
while the dashed and dot-dash lines are the contributions from the halo
and disk respectively.}
\label{vc_r_mh}
\end{figure}

The massive halo (MH) model had a Kuz'min-Toomre disk with $Q = 0.05$ initially and a halo polytrope index $m = {{3}\over{2}}$.  
The rotation curve for the MH model, shown in Figure \ref{vc_r_mh}, peaks at $R = 2.7$ with $\eta = 1.1$.  
Thus the dynamics of the disk are strongly influenced by the halo.

\begin{figure}[ht]
\plotfiddle{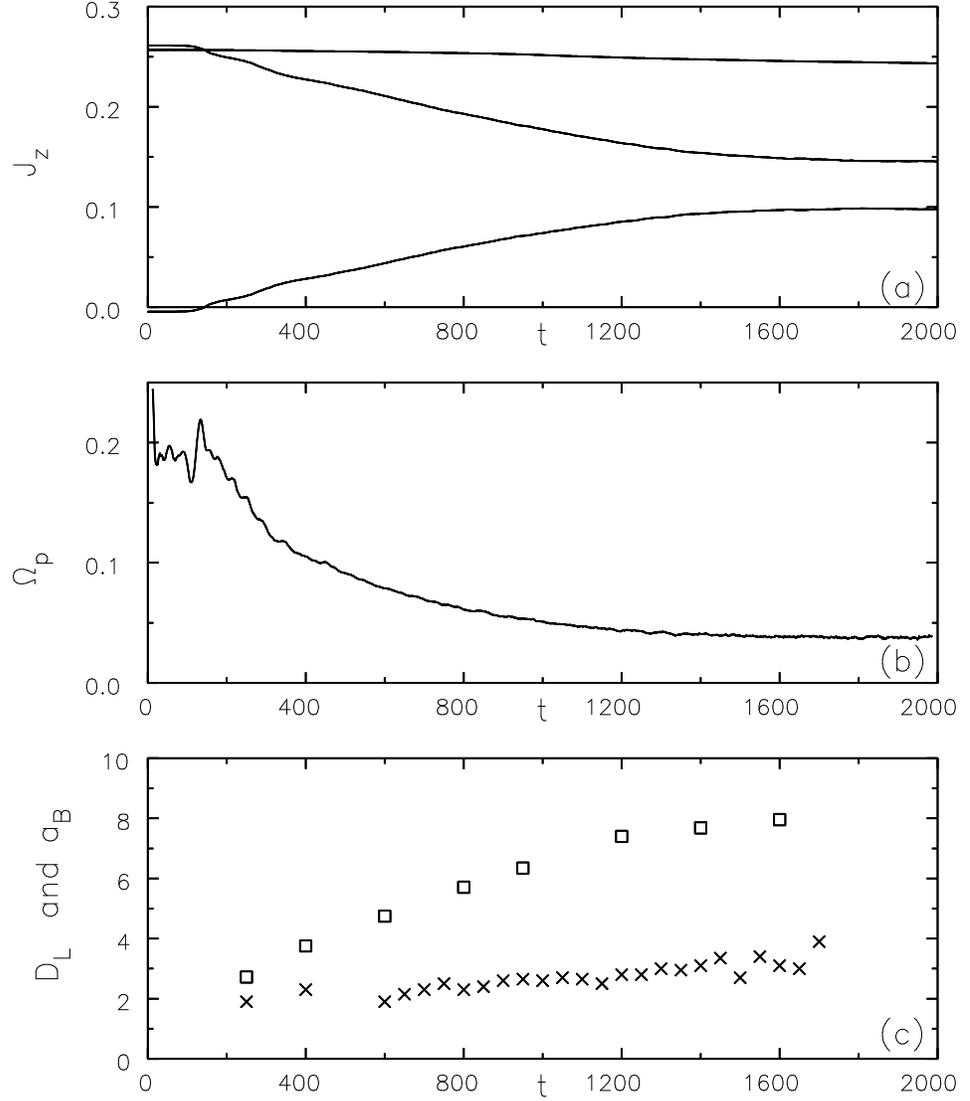}{6.0in}{0}{100}{100}{-290}{-170}
\caption{Evolution of the MH model.  (a) Angular momenta as
a function of time.  From top to bottom, the curves are total, disk
and halo angular momentum.  Note that angular momentum is being
transferred from disk to halo.  The slight drop in total angular
momentum from start to finish can be attributed to particle loss off
the grid.  (b) $\Omega_p$ as a function of time.  Note the rapid drop
in $\Omega_p$ after bar formation around $t = 100$.  (c) $a_B$
(crosses) and $D_L$ (squares) as functions of time.  At the last
measured point, $s \simeq 2.6$}
\label{jz_om_s_t_mh}
\end{figure}

By $t = 100$, the MH model forms a fast bar (in these units, the orbital period at $R = 2$ is $31$).  
This bar has been compared to that in NGC 936 (kinematic data of Kormendy, 1983).  
We found our bar was little more than a factor of two times as strong as the bar in NGC 936.  
With the formation of a bar, angular momentum starts to be transferred from the disk to the halo (Figure \ref{jz_om_s_t_mh}(a)).  
During this time, a trailing, bisymmetric wake can be identified in
the halo.  The trail angle is initially $\sim 45^\circ$, decreasing towards $0 ^\circ$ by $t = 1600$ when the net torque on the bar becomes negligible.  
The loss of angular momentum from the bar results in a significant drop in $\Omega_p$ (Figure \ref{jz_om_s_t_mh}(b)).  
Although the bar starts out with $ s \simeq 1.0 $, by simulation's end $ s $ has risen to $\sim 2.6$.  
This evolution, shown in Figure \ref{jz_om_s_t_mh}(c), occurs despite the continued growth of the bar throughout the simulation.

Other simulations with initial $Q = 1.0$ and $Q = 1.5$ gave similar results.  
We have also checked that halo rotation (both direct and retrograde) does not change our basic result, which is that a massive halo model cannot support a fast bar for a Hubble time.

\subsection{``Near maximum disk'' model}

\begin{figure}[ht]
\plotfiddle{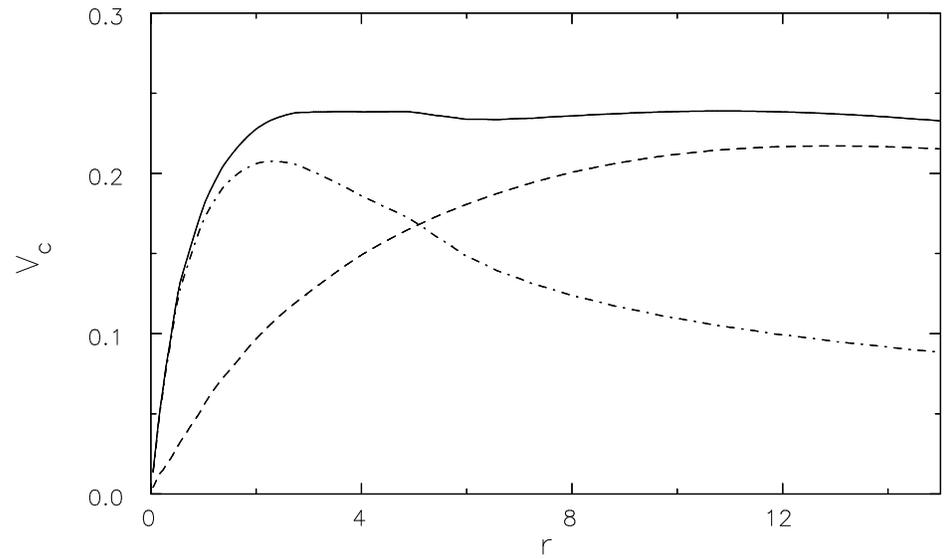}{4.0in}{0}{100}{100}{-290}{-270}
\caption{The circular velocity and its disk-halo decomposition for the
NMD model.  The continuous line is the total rotation curve,
while the dashed and dot-dash lines are the contributions from the halo
and disk respectively.}
\label{vc_r_md}
\end{figure}

We now report another model in which the disk dominated the inner scale lengths.  
The halo polytrope index was set to $m = {{1}\over{2}}$, the lowest value possible for stability against axisymmetric and radial instabilities.  
As can be seen from Figure \ref{vc_r_md}, at $R = 3.6$ where $v_c$ peaks, $\eta = 1.93$.  
It is in this sense that we refer to this as a ``near maximum disk'' (NMD) model: for a polytropic distribution function, the disk cannot be made any more dominant without making the halo unstable or unphysically hollow.  
The initial disk of this simulation was an exponential disk with $Q = 1.0$, resulting in a rotation curve which is more or less flat out to $R \simeq 15.$

\begin{figure}[ht]
\plotfiddle{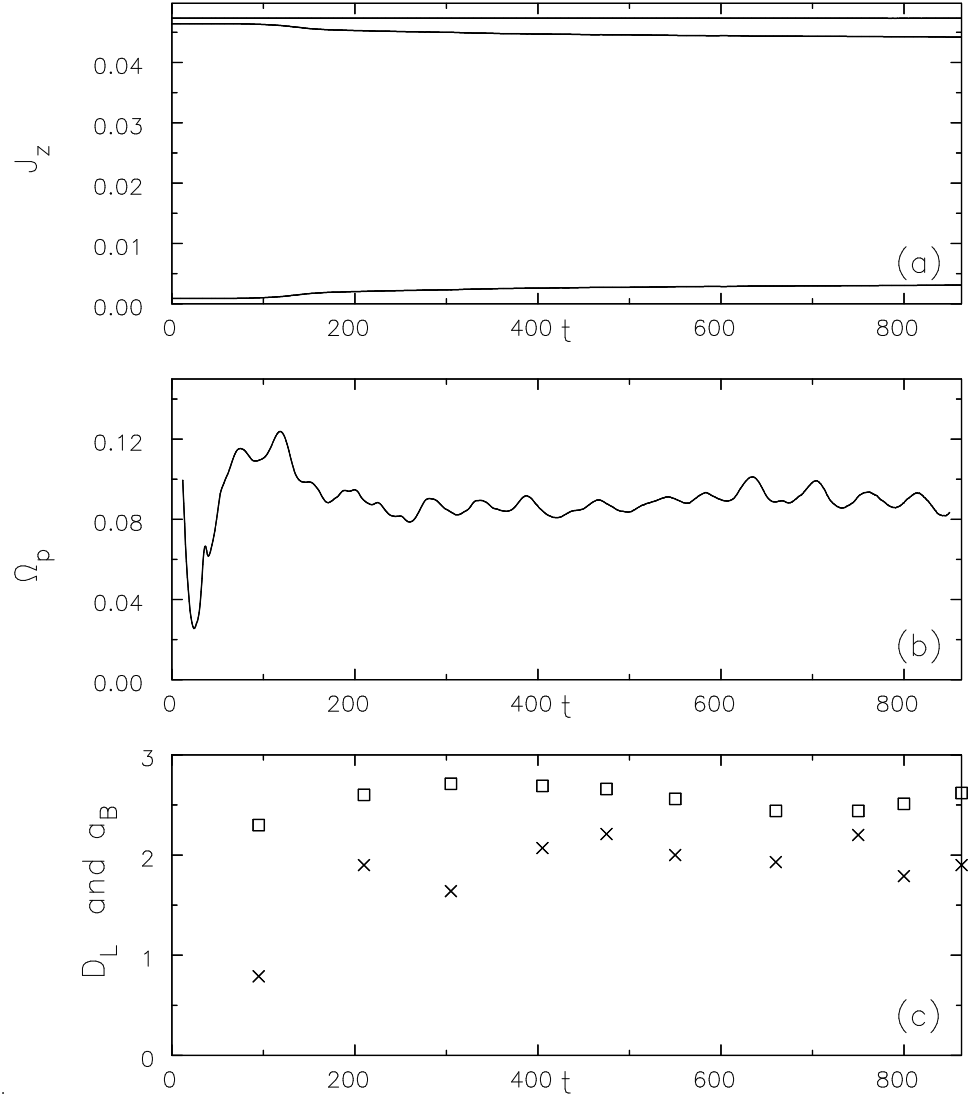}{6.0in}{0}{100}{100}{-290}{-170}
\caption{Evolution of the NMD model.  (a) Angular momenta as a
function of time.  From top to bottom, the curves are total, disk
and halo angular momentum.  Angular momentum transfer is much
diminished relative to the MH model.  (b) $\Omega_p$ as a
function of time.  Note the initial small drop in $\Omega_p$ after bar
formation at $t = 120$, followed by a much longer period when
$\Omega_p$ is stable.  The oscillations are caused by interference
with spirals.  (c) $a_B$
(crosses) and $D_L$ (squares) as functions of time.  At simulation's
end, $s \simeq 1.4$, i.e. the bar remains fast.}
\label{jz_om_s_t_md}
\end{figure}

The disk formed a fast bar by $t = 150$ (at $R = 2$, the orbital period is $55$ at $R=2$).  
Figure \ref{jz_om_s_t_md} shows the evolution of this NMD model.  
Although angular momentum is still being transferred from the disk to the halo, it is clear that the torque is much weaker in this case than in the massive halo model.  
The effect of this reduced torque is that the bar remains rapidly rotating for a Hubble time.  
At simulation's end ($t \simeq 850 $), $s$ has reached a value of $1.4$ and seems to be holding steady at that level.

It therefore seems likely that dynamical friction is not excessively strong in NMD models.  
Our result is still rather preliminary however, since so far we have only a single simulation in this regime.

\section{Discussion}

Assuming that all bars are fast, we have shown that barred
galaxies cannot be halo dominated within the inner few scale lengths.
De Blok et al. (these proceedings) find that low surface brightness (LSB)
galaxies are halo dominated at all radii.  The detection of slow bars in LSB
galaxies would then provide independent confirmation of this, while
presenting us with examples of previously unobserved slow bars.

We have also shown that near maximum disk models are free of the
dynamical friction problem, although this is still a preliminary
result to be explored further in future work.  If this result
turns out to hold in general, then continuity of bar strength from SB
to SAB to SA galaxies could perhaps be used as an argument in favor of near
maximum disks for all bright galaxies.

\end{document}